\documentclass[aps,pre,twocolumn,superscriptaddress,showpacs]{revtex4}
\usepackage{placeins}
\usepackage{graphicx}
\usepackage{amsmath}
\usepackage{times}
\usepackage{marvosym}

\begin{document}
\title{Stochastic Dynamics of Invasion and Fixation}
\author{Arne Traulsen}
\affiliation{Program for Evolutionary Dynamics, Harvard University, Cambridge MA 02138, USA}
\author{Martin A.\ Nowak}
\affiliation{Program for Evolutionary Dynamics, Harvard University, Cambridge MA 02138, USA}
\affiliation{Department of Organismic and Evolutionary Biology, Department of Mathematics, Harvard University, Cambridge, MA 02138, USA}
\author{Jorge M.\ Pacheco}
\affiliation{Program for Evolutionary Dynamics, Harvard University, Cambridge MA 02138, USA}
\affiliation{Centro de F{\'\i}sica Te{\'o}rica e Computacional, 
             Departamento de F{\'\i}sica da Faculdade de Ci{\^e}ncias, 
             P-1649-003 Lisboa Codex, Portugal}

\date{17 July 2006, Physical Review E 74, 011909}

\begin{abstract}
We study evolutionary game dynamics in finite populations. We analyze an evolutionary process, which we call {\it pairwise comparison}, for which we adopt the ubiquitous Fermi distribution function from statistical mechanics. The inverse temperature in this process controls the intensity of selection, leading to a unified framework  for evolutionary dynamics at all intensities of selection, from random drift to imitation dynamics. 
We derive, for the first time, a simple closed formula which determines the feasibility of cooperation in finite populations, whenever cooperation is modeled in terms of any symmetric two-person game. In contrast with previous results, the present formula is valid at all intensities of selection and for any initial condition. We investigate the evolutionary dynamics of cooperators in finite populations, and 
study the interplay between intensity of selection and the remnants of interior fixed points in infinite populations, as a function of a given initial number of cooperators, showing how this interplay strongly affects the approach to fixation of a given trait in finite populations, leading to counter-intuitive results at different intensities of selection. 
\end{abstract}

\maketitle

More than thirty years have passed since John Maynard-Smith laid the foundations of Evolutionary Game Theory \cite{maynard-smith:1973to}. 
In its context, the replicator dynamics equation \cite{taylor:1978wv,maynard-smith:1982to,hofbauer:1998mm} has inspired mathematicians, physicists, biologists, economists, psychologists, as well as political and computer scientists to investigate the evolution of traits under natural selection \cite{blume:1993aa, szabo:1998aa,nowak:2004aa}.  
In the absence of mutations, and whenever populations are infinite, homogeneous and well-mixed 
(each individual is equally likely to interact with any other individual), the (mean-field) description by the deterministic replicator dynamics in which the abundance $x$ of a strategy  changes according to its relative payoff, $\dot x = x (\pi_x - \langle \pi \rangle)$, has led to inumerous insights.   

In spite of the insights provided by deterministic replicator dynamics, it has been long recognized that populations, being finite, will exhibit stochastic effects which may play a sizable role. 
Such behaviour is well-known in physics, where stochastic effects in finite sized systems have been studied for a long time \cite{gardiner:1985bv,kampen:1997xg}.
Studies of stochastic evolutionary dynamics in finite populations, however, have been carried out mostly numerically.
Indeed, finite-size populations are an ever-present ingredient in individual-based computer 
simulations \cite{nowak:1992aa,nowak:1994aa,killingback:1996aa,hauert:2003aa,ifti:2004aa,szabo:2004aa,vukov:2005aa,santos:2005aa,hauert:2005mm}, which naturally incorporate such stochastic effects. 
Different intensities of selection have been employed in these studies, ranging from a strong selection framework, captured by the finite-population analogue of replicator dynamics \cite{killingback:1996aa,ifti:2004aa,szabo:2004aa,vukov:2005aa,santos:2005aa} to an extreme selection pressure under imitation dynamics, used as a metaphor of cultural evolution \cite{hofbauer:1998mm}. 

Recently, the fixation probability of a trait under frequency dependent selection in a finite population has been derived \cite{nowak:2004pw}. In order to control the interplay of drift and selection, an intensity of selection has been proposed, which measures the impact of the game on the fitness and determines the amplitude of fitness differences on which selection acts.
In such a system, fitness is not the number of offspring, but a measure for the potential to reproduce. 
Analytical results were obtained for weak selection, when the result of the 
game acts as a linear perturbation to neutral drift. The biological relevance of this weak selection stems from the fact that most evolutionary changes are near neutral, i.e.\ have small influence on fitness \cite{ohta:2002aa}. Therefore, neutral selection can serve as a reference point \cite{nowak:2004pw, antal:2005aa}. 
It has been shown that the finiteness of populations may indeed lead to fundamental changes in the conventional picture emerging from deterministic replicator dynamics in infinite populations. 
For instance, a single cooperator in a Prisoner's Dilemma has now a small, yet non-zero probability of survival in a finite population of defectors. Conversely, it is not always certain that defectors wipe out cooperators in finite populations \cite{taylor:2004aa,imhof:2006aa}. 

In physics, stochastic effects are often described in terms of an effective temperature, 
a methodology which has proven both successful and insightful at all scales. 
In line with this long tradition, we introduce here a temperature of selection in the evolutionary game dynamics of 
finite populations. Such temperature controls the balance between selection and random drift in finite 
populations, providing a convenient framework to study evolutionary game dynamics at all intensities of selection - from neutral, random drift, 
up to the extreme limit of cultural imitation dynamics. 
We define the {\it pairwise comparison} process, which utilizes the ubiquitous Fermi distribution 
function at finite temperature. Making use of the Fermi distribution, we provide for the first time 
a closed analytical expression for the fixation probability in a finite, well-mixed population of arbitrary size, for an arbitrary number of cooperators at start, and for an arbitrary intensity of (frequency dependent) selection. As a result, we are able to explore the role of the temperature of selection in evolutionary game dynamics in what concerns the overall fixation probabilities for different dilemmas.

General symmetric two player games can be interpreted in terms of social dilemmas of 
cooperation \cite{macy:2002aa,skyrms:2003aa,santos:2006aa,hauert:2006aa}, 
in which cooperators and defectors interact via the payoff matrix
\begin{equation}
\bordermatrix{
  & C & D \cr
C & R & S \cr 
D & T & P \cr}. 
\end{equation}
In all dilemmas mutual cooperation is favored ($R>P$), although individual reasoning often 
leads to mutual defection. 
In the Prisoner's Dilemma ($T > R > P > S$), the only stable evolutionary equilibrium of the replicator dynamics is the situation in which all individuals are defectors.
Other dilemmas of cooperation are more favorable to cooperation, though. For instance, 
whenever $T > R > S > P$, we enter the realm of the Snowdrift Game, where it is best to do the opposite of what the other player does: to cooperate if the other player defects and to defect if the other player cooperates \cite{hauert:2004bo}. This leads to the stable coexistence of cooperators and defectors under replicator dynamics, at the cooperator frequency
\begin{equation}
x^{\ast}=\frac{S-P}{T+S-R-P}.
\label{FP}
\end{equation}
In other words, starting from any given non-zero frequency of cooperators in the population, the system will evolve towards this coexistence equilibrium state. 
A third dilemma arises whenever $R > T > P > S$, the so-called Stag-Hunt game. 
In this case, one is better off doing whatever the other player will do, which corresponds to a coordination 
game favoring mutual cooperation ($R$). Indeed, replicator dynamics predicts the existence of an {\em unstable} interior fixed point 
at $x^{\ast}$ given by Eq.~(\ref{FP}), such that for frequencies above this point the population will evolve towards 
100$\%$ cooperation, the other fate being associated with startup frequencies below the unstable fixed point.

As usual in Evolutionary Game Theory, the fitness of each individual is associated 
with the payoff resulting from interactions with other individuals from the population. 
Under strong selection, an individual with higher fitness will always replace an individual
with lower fitness. 
Let us consider the following process : 
Two individuals are chosen randomly from the entire population - 
for example a cooperator $C$ and a defector $D$, with payoffs
$\pi_C$ and $\pi_D$, respectively. 
Under imitation dynamics, the fitter 
replaces the less fit with probability $p=1$. 
Similarly, under replicator dynamics selection always increases the fraction of the fittest type. 
However, most evolutionary changes have a small impact on fitness and sometimes the less successful individual may actually replace the more successful one. 
Whereas such occasional errors may be safely neglected in large populations, they
can have decisive effects in finite populations \cite{nowak:2004pw}. These effects are nicely captured by replacing the probability $p$ above with the Fermi function \cite{blume:1993aa,szabo:1998aa,nowak:2004pw,hauert:2005mm},
 \[
p= \frac{1}{1+e^{- \beta (\pi_C-\pi_D) }}, 
 \]
which gives the probability that $C$ replaces $D$. Such a process we call Pairwise Comparison. 
The inverse temperature $\beta \geq 0$ controls here the intensity of selection. For $\beta \to \infty$, the situation described above is recovered, 
i.e.\ the probability for replacement of the less fit individual is one. However, for any finite $\beta$, 
the reverse process may actually take place with a non-vanishing probability given by $1-p$. 
In particular, it is interesting to consider this process for $\beta \ll 1$, in which
the fitness has a marginal influence on replacement. For $\beta \ll 1$, the process described above 
reduces {\it exactly} to the frequency dependent Moran process under weak selection, studied in \cite{nowak:2004pw}.
\par
Let us consider the general payoff matrix Eq.~(\ref{payoffmatrix}).
When the number of interactions per individual is very high, the payoffs will depend only on the fraction of both types in the population. 
If there are $j$ cooperators and $N-j$ defectors, they will accumulate payoffs $\pi_C = (j-1) R + (N-j) S$  and  $\pi_D = j T + (N-j-1) P$, respectively \cite{rescaling}. Self interactions are excluded. 
Under Pairwise Comparison, and in the absence of mutations, only when the two individuals chosen have different strategies the total number of individuals 
with a given strategy can change by one.  
Similarly to the Moran process described in \cite{nowak:2004pw}, this defines a 
finite state Markov process with an associated tri-diagonal transition matrix.
The probabilities to increase the number of cooperators from $j$ to $j+1$, $T^+_j$,
and the probability to decrease that number from $j$ to $j-1$, $T^-_j$ are 
\begin{equation}
T^{\pm}_j = \frac{j}{N} \frac{N-j}{N} \frac{1}{1+e^{\mp  \beta (\pi_C-\pi_D)}} .
\end{equation}
For large populations, this process can be approximated by a stochastic differential equation with drift $T^{+}_j-T^{-}_j$ and diffusion $\sqrt{(T^{+}_j+T^{-}_j)/N}$  \cite{traulsen:2005hp}.  For the pairwise comparison process, this yields
\begin{equation}
\dot x = x(1-x) \tanh \left[ \frac{\beta}{2} \left(\pi_C-\pi_D \right) \right] + \sqrt{\frac{x(1-x)}{N}} \xi
\end{equation}
where $x=i/N$ is the fraction of cooperators and $\xi$ is Gaussian white noise with variance one. For $N \to \infty$, the stochastic term vanishes. In this limit, the replicator dynamics \cite{hofbauer:1998mm} is recovered from the first order term of a high temperature expansion, $\beta \ll 1$.

In finite populations, the quantity of interest in this process is the fixation probability of cooperators, i.e.\ the probability to end up in the state with $N$ cooperators given that the initial number is $k$. This probability depends only on the ratio $\alpha_j = T^-_j / T^+_j$. For the Pairwise Comparison process, this ratio reduces to $\alpha_j = \exp \left[{-  \beta (\pi_C-\pi_D)}\right] $. 
In terms of $\alpha_j$, the fixation probability $\phi_k$ is given by \cite{karlin:1975xg}
 \begin{equation}
\phi_k = \frac{\sum_{i=0}^{k-1} \prod_{j=1}^{i} \alpha_j}{
\sum_{i=0}^{N-1} \prod_{j=1}^i  \alpha_j},
\label{fixprobMoran1}
 \end{equation}
and the payoff difference can be written as $\pi_C-\pi_D= 2  u j + 2v $,
where $2u = R-S-T+P$ and $2 v = -R+S N-P N+P$, which simplifies
the fixation probability to 
 \begin{equation}
\phi_k= \frac{\sum_{i=0}^{k-1} \exp \left[- \beta i (i+1)  u - 2 \beta i v  \right]
}{
\sum_{i=0}^{N-1} \exp \left[-\beta i (i+1)  u - 2 \beta i v  \right]} \; . 
 \end{equation}
One can interpret the sums in the previous equation as numerical approximations to the integral 
$\int_{-\frac{1}{2}}^{k-\frac{1}{2}}  \prod_{j=1}^{i} \alpha_j di$. The error of this approximation is of order $N^{-2}$ and will therefore rapidly vanish with increasing $N$.  
Replacing the sums by the integrals leads to the surprisingly simple formula, 
\begin{equation}
\phi_k= \frac{{\rm erf}\left[ \xi_k\right]-
{\rm erf}\left[ \xi_0\right]}
{{\rm erf}\left[\xi_N \right]-
{\rm erf}\left[ \xi_0\right]}
\label{fixation}
\end{equation}
valid for $ u \neq 0$ where 
${\rm erf}(x)=\frac{2}{\sqrt{\pi}}\int_0^x dy\, e^{-y^2}$ is the error function
and 
$\xi_k=\sqrt{\frac{\beta}{ u}} \left(k  u + v \right) $. 
For $u=0$, which is in Game Theory sometimes referred to as "equal gains from shifting", we find instead
\begin{equation}
\phi_k = \frac{e^{-2 \beta v k}  -1}{e^{-2 \beta v N}  -1}.
 \label{pd}
 \end{equation}
This result is formally identical to the fixation probability of $k$ individuals with fixed relative fitness $r=e^{2 \beta v}$ \cite{kimura:1983aa,ewens:2004qe}. Eqs.\ (\ref{fixation}) and (\ref{pd}) reduce trivially to the neutral selection result $\phi_k =k/N$ for $\beta \to 0$. 
In contrast with the results from Ref.~\cite{nowak:2004pw}, these closed expressions are valid for {\em any} intensity of selection and for any given initial number of individuals of a given trait. 
Two limiting cases follow immediately: $\beta \ll 1$ corresponds to weak selection and yields the generalization of the 
weak selection result for the birth-death process given in \cite{nowak:2004pw} to an arbitrary initial number of mutants. 
Strong selection is described by $\beta \gg 1$ and reduces the process to a semi-deterministic imitation process: The speed of this process remains stochastic, but the direction always increases individual fitness for $\beta \to \infty$. This limit, which corresponds to the 
cultural evolution models of imitation dynamics, is outside the realm of the birth-death process studied in \cite{nowak:2004pw} 
and results from the nonlinearity of the Fermi distribution function.
\par
In the following we study the evolutionary game dynamics of finite populations in some concrete cases, in connection with 
the results shown in Fig.~1.  

\begin{figure}[hpbt]
\begin{center}
{\includegraphics[width=8.5cm]{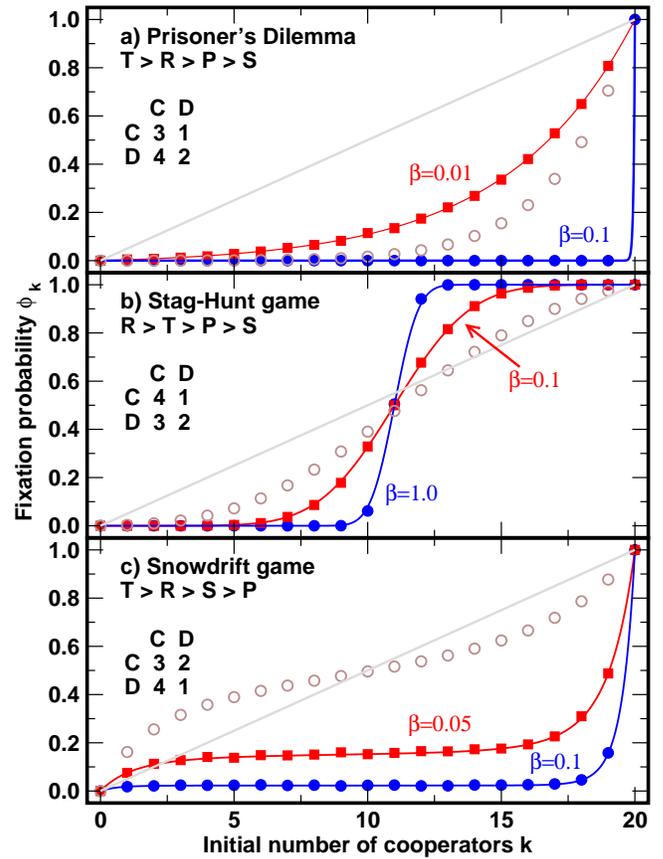}}
\caption{Fixation probabilities for the three qualitatively different dilemmas of cooperation 
describable by $2 \times 2$ games. 
Simulation results (filled symbols)
coincide perfectly with the theoretical result Eqs.~(\ref{fixation}) and (\ref{pd}) (solid lines). For comparison, the theoretical result for the Moran process at maximum selection intensity ($w=1$, open circles) and the neutral selection result ($\beta=0$, diagonal line) are shown. 
(a) In the Prisoner's Dilemma, defection is dominant. As a result, only for very weak selection ($\beta = 0.01$) cooperators acquire significant chances of fixation. 
(b) The Stag-Hunt game has an unstable equilibrium at $k^{\ast}=(N+2)/2$, and the transition across this point depends sensitively on the intensity of selection.
(c) The Snowdrift Game exhibits an internal equilibrium at $k^{\ast}=(N-2)/2$, which is an inflection point in the theoretical curves. 
A small population size of $N=20$ has been chosen on purpose to stress the quality of the theoretical result 
related to Eqs.~(\ref{fixation}) and (\ref{pd}).  Each simulation result corresponds to the fraction of fixation of cooperators in $10^4$ independent realizations.}
\end{center}
\end{figure}

For the Prisoner's Dilemma 
we choose $R=3$, $S=1$, $T=4$, and $P=2$. Since $u=0$, the fixation probability is given by Eq.~(\ref{pd}). In Fig.~1a we show a comparison between the analytical results and extensive computer simulations for a small population of $N=20$ individuals. Clearly, cooperators are always in disadvantage with respect to defectors. In spite of the exponential increase of the fixation  probability of cooperators with the initial number $k$, only for very weak selection ($\beta = 0.01$) do cooperators acquire reasonable chances in a population as small as $N=20$. As stated before, 
the error of the approximation leading to Eqs. (\ref{fixation}) and ~(\ref{pd}) is of order $N^{-2}$. 
However, even for $N=20$ excellent agreement with numerical simulations is obtained as illustrated in Fig.~1. 

As a second numerical example, we consider the Stag-Hunt game given by the ranking 
of the payoff matrix illustrated in Fig.~1b (note that we only changed the order of numerical values from Fig.~1a). 
This system exhibits two pure Nash equilibria, i.e.\ each strategy is the best reply to itself, 
and any selection mechanism leads to a bistable outcome. The replicator dynamics predicts a sharp transition from full-defection to full-cooperation at the unstable fixed point $x^{\ast}$ given by Eq.~(\ref{FP}).  
In finite populations, this point is given by $k^{\ast}=N x^{\ast} + (P-R)/(T+S-R-P)$, where the second term resulting from the exclusion of self interactions is irrelevant for large $N$. 
The transition at $k^{\ast}$ depends sensitively on the intensity of selection, becoming sharper the larger the 
intensity of selection. For the Moran process discussed in \cite{nowak:2004pw} 
a smooth transition from zero to one is always observed in all dilemmas,
even when the intensity of selection is maximized ($w=1$), for which the fitness equals the payoff resulting from the game 
(open-circles in Fig.~1). Hence, only for very smooth transitions between the two absorbing states 
do the predictions in \cite{nowak:2004pw} apply. Eq.~(\ref{fixation}), instead, provides the overall scenario 
for all intensities of selection without need to invoke the so-called $1/3$-rule. 

The Snowdrift Game \cite{hauert:2004bo} (also known as 'Hawk-Dove' game or 'Chicken') characterized by the payoff ranking $T>R>S>P$ presents the most interesting scenario in this context. This system exhibits a mixed Nash equilibrium at $k^{\ast}$. 
Replicator dynamics predicts that any initial condition will lead to this mixed-equilibrium. 
This is clearly not possible in a finite population, in which 
the system ends up in one of the 
absorbing states $k=0$ or $k=N$. 
For weak selection, the fixation probability increases 
smoothly with $k$ from zero to one 
(open circles in Fig.~1c). 
As $\beta$ increases, a plateau develops in the center region from $25 \%$ up to $75 \%$ of cooperators, in which the 
fixation probability is roughly independent of $k$. 
For $\beta=0.05$ its value 
is $\approx 15 \%$. For $\beta=0.1$, the plateau widens, but more interestingly, it shifts position, so that now the fixation probability 
becomes $\approx 2 \%$. 
This behavior 
is related to the location of 
$k^{\ast}$, which in the finite system 
leads to the occurrence of an inflection point at $k^{\ast}$ in the fixation probability curves. 
With increasing selection pressure the 
probability distribution becomes sharply peaked 
around $k^{\ast}$.
Hence, the fixation probability of {\it defectors} rapidly approaches one, as $k^{\ast}<1/2$. In this case, the remnant of the interior fixed point leads to a fast increase of the time to fixation, which may become arbitrarily long (a detailed account of the fixation times will be published elsewhere \cite{TPN:2006aa}). This observation cannot be inferred from the weak selection analysis of Ref.~\cite{nowak:2004pw} and contrasts sharply with the predictions from the replicator dynamics equation: For the stochastic system, fixation in $x=0$ or $x=1$ is certain, whereas the replicator dynamics predicts coexistence of cooperators and defectors.

Throughout this work, we have followed the conventional approach of evolutionary game theory assuming 
that the payoff matrices are constant and independent of $\beta$. 
However, the intensity of selection and/or 
the payoff matrix may change during evolution, a feature which would lead to a 
coevolutionary process 
that deserves further attention. 
One possibility is a situation in which there is a certain 
amplitude of variation in the payoff values during each interaction, for fixed $\beta$. 
To the extent that these variations can be considered random, 
such an effect might contribute (in a mean-field treatment) to an overall rescaling of the ``selection temperature'' leading to a new value 
$\beta'$.
Such new temperature would result from replacing the distribution of fitness values among each of the two types in the population by their average values. 
The theory presented here may be viewed as accounting for such a mean-field description, which will become more 
exact the narrower the payoff distributions. 

To sum up, the pairwise comparison process discussed here provides an unified framework in which one can study evolutionary 
dynamics in finite populations. We derived a simple formula which determines the fixation probability for a given trait under frequency dependent selection, 
at all intensities of selection and for any given number of initial individuals of that trait present in the population. 
In a finite population, the only stable evolutionary outcome corresponds to all individuals sharing a common trait. The probability of fixation in either of these traits, however, is a non-trivial function of the number of individuals sharing that trait at start, depending sensitively 
on the intensity of selection, thereby configuring an interesting evolutionary scenario not accountable for previously. 
We have shown that with increasing selection pressure the topological structure which is predicted by the deterministic replicator dynamics 
in infinite populations plays an increasing role. Depending on the nature of the game, we find that the transition between competing extrema in 
coordination games exhibiting bistability gets sharper with increasing selection pressure. On the other hand, for games 
exhibiting a mixed equilibrium, the fingerprint of the stable mixed equilibria reflects itself in the rate of fixation into one of the absorbing states, which may increase prohibitively with increasing selection pressure. In other words, even when fixation 
is certain, the time to reach it may become arbitrarily long. 

\acknowledgments{
We thank two anonymous referees for their suggestions.
Discussions with C.\ Hauert, H.\ Ohtsuki and C.\ Taylor are gratefully acknowledged. 
A.T.\ acknowledges support by the ``Deutsche Akademie der Naturforscher Leopoldina'' (Grant No.\ BMBF-LPD 9901/8-134).
J.M.P.\ acknowledges financial support from FCT, Portugal. 
The Program for Evolutionary Dynamics at Harvard University is sponsored by Jeffrey Epstein. 
}

\bibliographystyle{plain}

\end{document}